\documentclass[onecolumn,preprint,preprintnumbers,amsmath,amssymb,prl,footinbib]{revtex4}
\usepackage{graphicx}
\usepackage{amssymb}
\usepackage{epstopdf}
\usepackage{natbib}
\usepackage{pifont}
\newcommand{\half}{\frac{1}{2}}
\newcommand{\ket}[1]{\vert{#1}\rangle}
\newcommand{\bra}[1]{\langle{#1}\vert}

\begin{document}
\bibliographystyle{apsrev}
\title{Time-keeping with electron spin states in diamond}
\author{J.S. Hodges$^{1}$}
\email{jsh58@columbia.edu}
\author{ D. Englund$^{1,2}$}
\email{englund@columbia.edu}
\affiliation{$^{1}$Dept. of Electrical Engineering}
\affiliation{$^{2}$Dept. of Applied Physics and Applied Mathematics, Columbia University, New York, New York 10027}
\begin{abstract}{Frequency standards based on atomic states, such as Rb or Cs vapors, or single trapped ions, are the most precise measures of time.
Here we introduce a complementary device based on spins in a solid-state system --- the nitrogen-vacancy defect in single crystal diamond.  We show that this system has comparable stability to portable atomic standards and is readily incorporable as a chip-scale device.  Using a pulsed spin-echo technique, we anticipate an Allan deviation of $\sigma_y = 10^{-12} \tau^{-1/2}$ with current photoluminescence detection methods and posit exceeding $10^{-14}$ with improved diamond material processing and nanophotonic engineering.}
\end{abstract}
\date{\today}                                           

\maketitle

Atomic clocks are the most accurate systems for measuring time and frequency. They are used in a broad range of applications, including communication, computing systems and navigation, such as the global positioning system (GPS).  Modern frequency standards derive their stability from the precisely measured internal hyperfine level splittings of atoms of Cs, Rb, or H.  When an oscillating magnetic field is resonant with the energy difference of these internal  states a change in population between levels changes the radiofrequency or optical absorption.  Standard lock-in techniques modulate the driving frequency and monitor the absorption as a correction for a tunable active reference oscillator, e.g. a quartz crystal, thus stabilizing it to the atomic line \cite{Camparo:RbClock:PhysToday}.  Recent experiments on single trapped ions \cite{RosenbandScience2008} and ensembles of atoms trapped in optical lattices \cite{TakamotoNature2005,LudlowScience2008} have far exceeded the international cesium standard, even enabling the observation of general relativity corrections within a few meters \cite{ChouScience2010}.  Such precision, however, comes at the expense of mobility, as the infrastructure for these standards encompass several tens of cubic meters of space.  At the other extreme, portable standards based on rubidium vapor cells provide excellent stability for time scales ranging from 1~s to 10$^{4}$~s and find usage in satellites, laboratory equipment, and cellular communications.  Mobile devices, which typically do not contain their own precision standards, can share GPS time signals for maintaining communication standards, but when the external lock signal is obstructed
a precise local frequency standard with minimal drift is necessary to maintain synchronization.

\begin{figure}[htbp]
\begin{center}
\includegraphics[scale=.85]{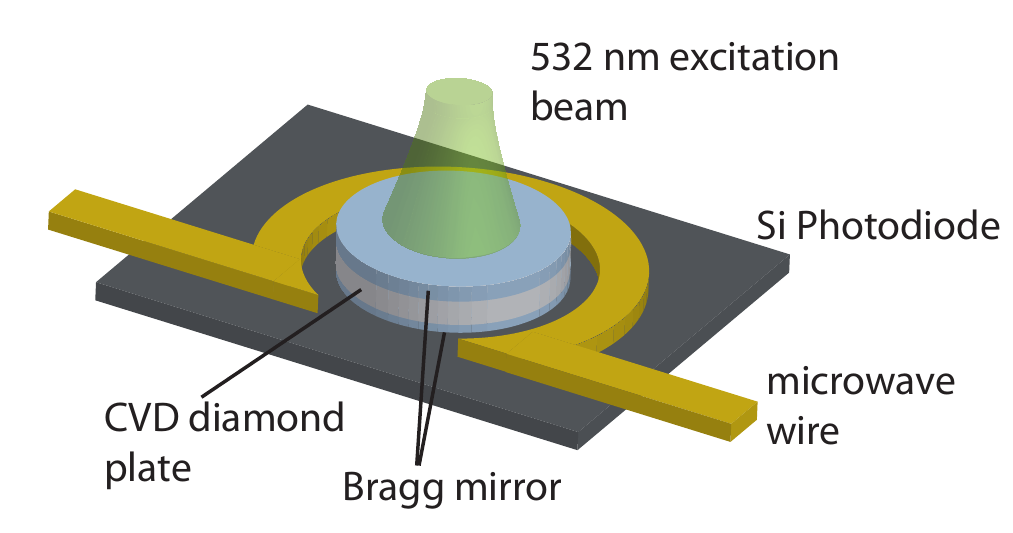}
\caption[Schematic of the diamond frequency standard.]{Schematic of the diamond frequency standard. A thin (100 $\mu$m) diamond chip is surrounded by dielectric stacks (Bragg Reflectors) on both sides to create a resonant cavity for 532nm excitation in order to reduce the power requirements.  On-chip 532nm excitation comes from a doubled 1064nm surface emitting laser (not shown).  Silicon photodetectors under the diamond collect emission from face.  Microwaves, which address the NV magnetic sublevels, are applied to the entire sample by a planar stripling.}
\label{fig4}
\end{center}
\end{figure}

To the address this need, several groups have miniaturized these atomic standards \textit{on-chip} through the aid of modern microfabrication techniques applied to detectors and lasers \cite{MorelandKitchingAPLClock,HoneywellChipClock, Shah:Nph:ChipMag}.  Here, we propose 
 a solid-state alternative based on electron spin states in the negatively charged nitrogen vacancy center (NVC) in diamond.  The diamond system offers a host of advantages.  First, single crystal diamond can be grown into a micron-scale, radiation hard chip, which makes it portable and well-suited for integration in a semiconductor fabrication process \cite{Toyli:2010zr}.  Second, this solid-state system derives its performance as a clock from exceptional spin lifetimes of the NVC \cite{Balasubramanian:NMat} and resembles atomic and molecular systems.  The optical detection of the NVC also increases the signal-to-noise of solid-state standards based on inductive detection \cite{WhiteSSS}.  Compared to vapor cells, this standard does not suffer from doppler or collisional broadening.  Although it exhibits an increased homogeneous linewidth compared to atomic standards, due to a complex mesoscopic environment \cite{Childress:Science}
 , a higher density of defects in solids allows for comparable frequency stability in smaller sensor volumes. We estimate a stability, expressed as an Allan deviation, of $\sigma_y \approx2\times10^{-12}$ at 1s of averaging for a 0.1 mm$^{3}$ diamond sample.  Paired with potentially modest operational power requirements, this NV clock could result in a new generation of portable solid-state frequency standards.  A proposed device is shown in Figure \ref{fig4} and contains: a diamond chip grown with a dielectric cavity for lowering the optical pumping power requirements, photoluminescence (PL) collection with on-chip Si photodiodes, and planar microwave waveguides for addressing hyperfine transitions, all of which can be integrated into modern device fabrication.
%



An atomic clock derives its stability from the large quality factor, $Q=\nu/\Delta \nu$, of the probed resonance, with narrow linewidth, $\Delta\nu$, being much smaller than the resonant frequency $\nu$.  In the solid-state, we desire: (i) a microwave transition ($\nu \sim$ GHz) that also exhibits a large Q; (ii) a resonance that does not vary due to material or fabrication processes; and (iii) a precise method of measuring population changes. The NVC in diamond satisfies these criteria having a ground state spin triplet with long ($>$1ms) spin coherence time (thus a small linewidth); a crystal field splitting of the ground state with an intrinsic resonance frequency near 2.8GHz that is independent, to lowest order, of applied magnetic field; and spin states that are optically polarizable and measurable on single molecule length scales, due to the spin-dependent relaxation of the defect.

The magneto-optical description of the NVC is well-documented in the literature, thus we give only a brief phenomenological summary.  Figure 2A shows the relevant spin sublevels (0,1,2,3 and S) for the NVC.  Optical absorption of green laser light causes for broadband PL of the NVC from 637-800 nm.  A spin-dependent intersystem crossing between excited state triplet (3) to a metastable, dark singlet level (S) changes the integrated PL for the spin states $\ket{0}$ and $\ket{\pm1}$.  The deshelving from the singlet occurs primarily to the $\ket{0}$ spin state, providing a means to polarize the NVC.  Microwave fields resonant between levels 0 \& 1, perturb the spin populations, and thus the PL response.  This response can be measured in a steady-state (or continuous wave) response to simultaneous optical and microwave fields (e.g. Figure 2B); or, in a pulsed manner, by preparing a state using only microwaves, and observing transient PL response.

In order to quantitatively compare both methods, we construct a simple model for the relevant spin dynamics described by Hamiltonians for the lowest and first excited triplet state and two metastable singlet states.  For the purpose of a frequency standard, we monitor the response of only the ground  state triplet sublevels to resonant excitation, yielding the ground state Hamiltonian\cite{Dolde:2011ys}:
\begin{eqnarray}
\mathbf{H}_{gs} = (D_{gs} + d_{\parallel} \sigma_z) S_z^2 + g \mu_b \vec{S}  \cdot \vec{B} + \nonumber \\
d_\perp \sigma_x (S_x S_y + S_y S_x) +  d_\perp \sigma_y (S_x^2 -S_y^2)
\end{eqnarray} 

\begin{figure}[htbp]
\begin{center}
\includegraphics[scale=.3]{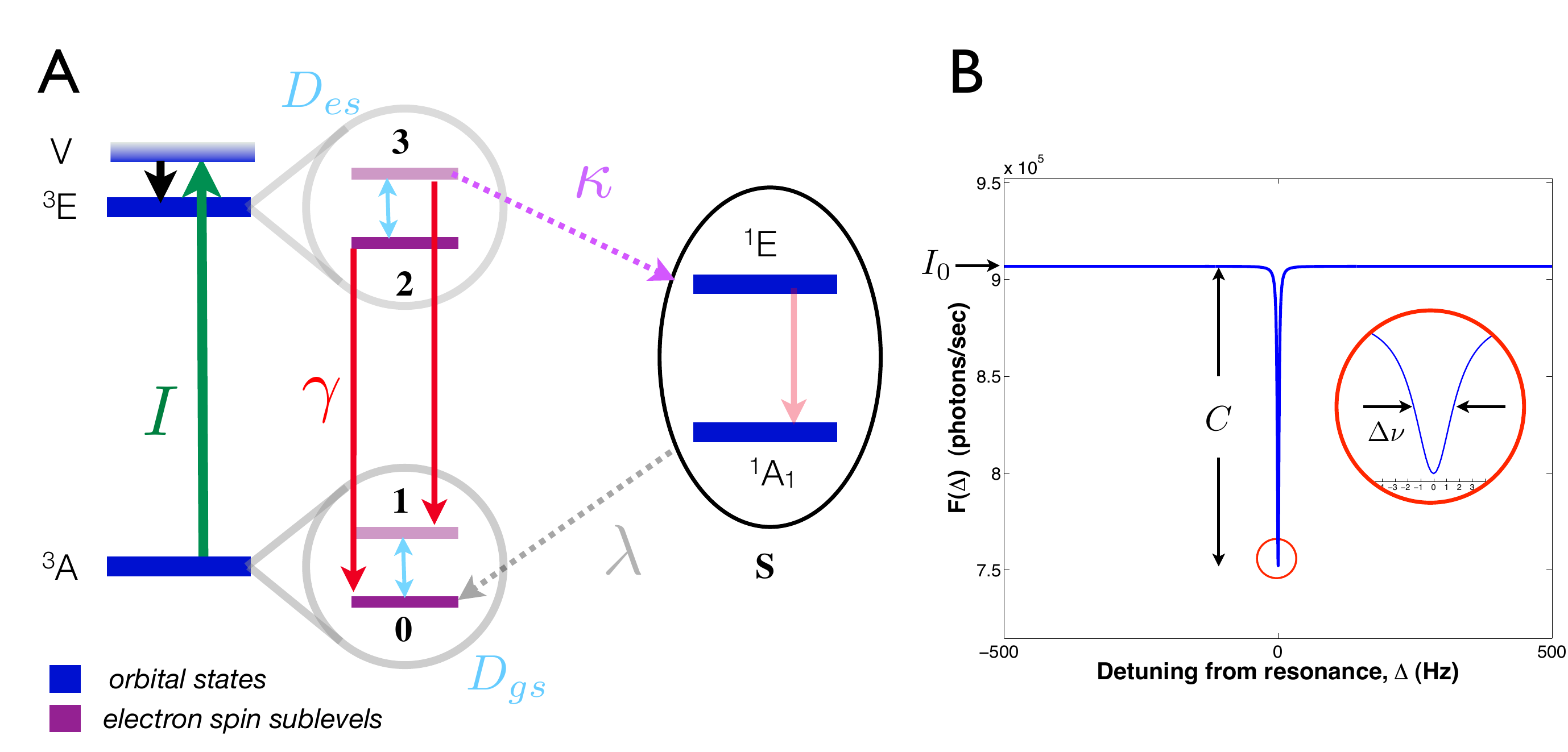}
\caption[Nitrogen-vacancy center energy levels and resonant response.]{Nitrogen-vacancy center energy levels and resonant response.  A.  The lowest lying triplet ($^3E$, $^3A$) and singlet ($^1E$, $^1A$) orbital states of the NV$^{-}$ center.  $I$, $\gamma$, $\kappa$, and $\lambda$ represent the absorption, PL, intersystem crossing, and deshelving rates, respectively.  Sublevels 0 and 2 are $S_z$ eigenstates $\ket{0}$, whereas 1 and 3 are $\ket{\pm1}$, with the degeneracy lifted by small crystal strain or applied magnetic field.  The model is simplified by approximating both singlet states as a single metastable level.  B.  The steady-state fluorescence emission of the NVC under continuous optical and microwave irradiation, detuned from resonance by $\Delta$.  The PL response function, as derived from a master equation approach, is approximated by a Lorentzian: $F(\Delta) = I_0 (1 - \frac{C\,\,\delta\nu^2}{(\Delta/\pi)^2 + \delta\nu^2})$, where $C$ is the modulation depth and $\delta\nu$ is the FWHM \cite{dreau2011}.}
\label{fig1}
\end{center}
\end{figure}

Here $d_{\parallel,\perp}$ are the ground state electric dipole moment components along and perpendicular two the $C_{3v}$ symmetry axis of the defect.  $D_{gs}$ is the ground state crystal field splitting (2.870 GHz), $\mu_b$ is the Bohr magneton, and g is the Lande factor (assumed to be isotropic).  $S_k$ are spin-1 operators in the $k=\{x,y,z\}$ directions.  The local electric field vector, induced by crystal strain, is $\vec{\mathbf{\sigma}}$.  In the limit of static magnetic and electric fields much smaller than $D_{gs}$, the eigenfunctions are those of the $S_z$ operator, as show in Fig. \ref{fig1}.

A driving field at frequency $\omega$ induces electron spin resonance (ESR)  transitions between $\ket{0}$ and $\ket{\pm1}$.  On resonance ($\omega \approx D_{gs}$), the PL decreases and provides a feedback signal as to lock $\omega$ to $D_{gs}$.  The dynamics are best viewed as a response to a time-varying magnetic field $B_1 = 2 b_1 \cos(2\pi\omega t) \hat{x}$ and transforming $H_{gs}$ into an interaction frame defined by the operator $V = e^{2\pi i \omega t S_z^2}$.  Under the rotating wave approximation, we have:
\begin{equation} 
\label{hint}
\mathbf{H}_{gs}^\prime =  (D_{gs} + d_{\parallel} \sigma_z - \omega) S_z^2 + g \mu_b B_z S_z + g \mu_b b_1 S_x
\end{equation}
 
The relaxation rates of the excited triplet and singlet states, shown in Fig. \ref{fig1}A, play an important role in the optical pumping and spin measurement \cite{Manson:2006p89}.  We model the total magneto-optical response using a master equation approach:
\begin{equation}
\dot{\rho}  = \frac{1}{i\hbar} [H_{gs}^\prime, \rho] + \sum_k L_k \rho L_k^\dagger - \half L_k^\dagger L_k \rho - \half \rho L_k^\dagger L_k
\end{equation}
where $\rho$ is the density operator for the NVC ground, excited triplet, and effective singlet states.  The jump operators, $L_k$, have magnitudes corresponding to relaxation rates $\sqrt{r_k}$.  Conveniently, the solution to the equation yields the total magneto-optical response for both continuous and pulsed excitation, allowing for numerically assessment of clock sensitivity with changes to electromagnetic fields.

Under continuous excitation of optical and microwave fields the NVC frequency standard closely resembles a two-isotope Rb standard.  Spin-dependent PL of the NVC occurs under non-resonant absorption of green light of intensity $I$.  Application of a microwave field of intensity $\Omega = g\mu_b b_1$, detuned from resonance by an amount $\Delta = D_{gs} - \omega$,  causes a broad, phonon-assisted PL: $F(I,\Omega,\Delta) = \gamma \rho_{22}^{ss} + \frac{\gamma^2}{\kappa + \gamma} \rho_{33}^{ss}$.  Here $\rho_{22}^{ss}$ and $\rho_{33}^{ss}$ represent the populations of the first excited state spin sublevels in the steady-state, as analytically derived from the master equation description of the system \footnote{\textit{A manuscript describing the details of this model is in preparation.  See \cite{dreau2011} for a similar approach.}}.  Figure \ref{fig1}B shows a typical response for $F$ for varied detunings, which displays a Lorentzian lineshape.  The stability of the clock can be derived from the resonance curve by considering the Allan variance:
\begin{equation}
\sigma_y (\tau) = \frac{1}{2\pi Q }\frac{1}{(S/N)} \frac{1}{\sqrt{\tau}}
\end{equation}
where $S/N$ is the signal-to-noise ratio, which depends on the photon shot noise and imperfect modulation of the resonance (i.e. $C\neq1$), and $\tau$ is the averaging time.  The intrinsic linewidth is limited by the paramagnetic and nuclear spin environments which fluctuate during the measurement.  The linewidth broadens if the microwave and optical transitions are driven near saturation (`power broadening'), but higher pump powers also increase the the depth of the dip ($C\rightarrow 1$).  Far below optical saturation, the PL rate is sufficiently small, and modulation depth (C) reduced, to cause a decrease in stability per averaging time.  As the Q is the most important factor for the stability, the optimal condition is attained for values which offset the broadening with the reduced signal.  Under these conditions we estimate a linewidth of 3.6 MHz, an off-resonance fluorescence rate of $\sim$9400 photon/s (accounting for finite detector efficiency), and a 17\% modulation depth. We thus obtain $\sigma_y(\tau) = 8.124\times10^{-5}\,\, \tau^{-1/2}$ for a \textit{single} NVC.

In short, the NVC lacks sufficient stability when monitoring the PL response continuously.  The laser excitation must be reduced far below saturation for the optical power broadening of the line to reach the homogeneous linewidth.  At such low pump powers, the fluorescent photon flux is so small that the gains in $Q$ are offset by losses in $S/N$. As seen in magnetometry studies of the NVC \cite{MazeMagnetometer}, a pulsed microwave excitation scheme that monitors transient fluorescence behavior can improve the performance over a continuous excitation/detection method.

Alternatively, the NVC crystal field splitting ($D_{gs}$) can be monitored in a pulsed fashion, akin to Ramsey spectroscopy used in atomic clocks.  The standard two-pulse Ramsey sequence, with time separation $T$, imprints a phase proportional to the frequency drift, $\delta\omega$, between two of the hyperfine states of the atomic system.  The PL response varies sinusoidally with $\delta\omega T$ and can be linearized to provide a passive standard with frequency uncertainty set by $~T^{-1}$.  However, if the system does not remain coherent for times $T >>T_c$, the modulation of the PL response proportional to the phase does not persist.  In atomic vapor cells, trapped ions, or atomic fountains, $T_c > $ 1s due to magnetic shielding and minimizing atomic collisions.  In the NV system, due to a complex solid-state environment consisting of many nuclear spins coupled to a single defect, $T_c$ is limited to tens of microseconds in a typical sample.  Quantum memories and magnetometers of AC fields employ a Hahn echo sequence to extend the coherence time to $T_2$ by removing slowly varying magnetic fields.  For frequency standards based on $m_f = 0$ ``clock'' states, the additional $\pi$-pulse of the echo sequence completely removes the phase accumulation of frequency drift, making it useless for time-keeping.  The S=1 nature of the NVC, however, contains symmetries within the quadrupolar moments that allow for an modified echo sequence to yield a PL signal proportional to the drift for $T ~ T_2$.

\begin{figure}[htbp]
\begin{center}
\includegraphics[scale=.30]{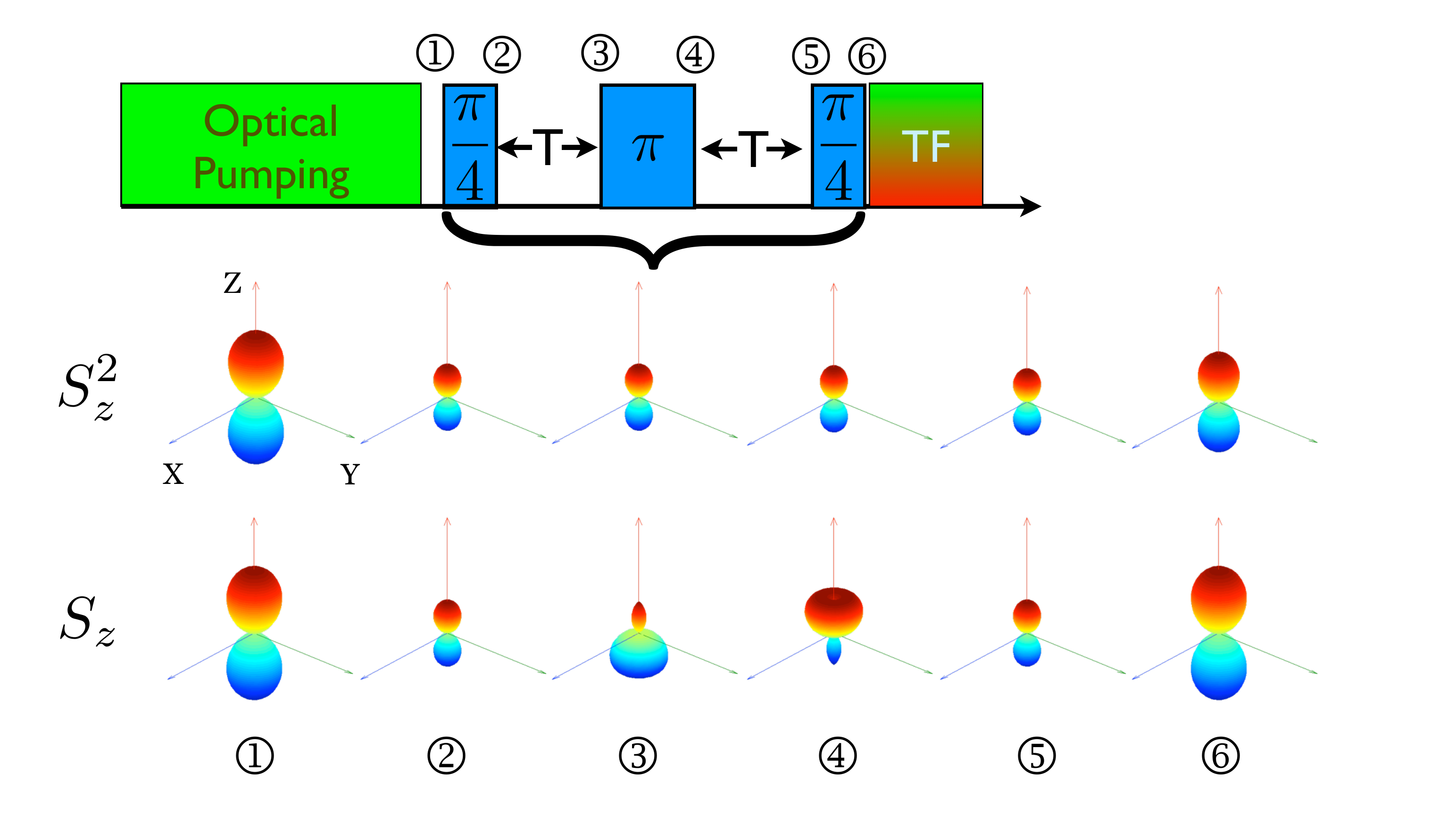}
\caption[Spin-1 time keeping pulse sequence]{Spin-1 time keeping pulse sequence.  The spin-1 ground state $\ket{0}$ state is prepared by optical pumping through the singlet state. 45$^\circ$ rotations ensure that the $S_z$ term cancels with the echo sequence, while the $S_z^2$ does not, as illustrated in the projected spherical harmonics (inset).  A transient fluorescence (TF) measurement records the photocurrent for about 300ns timed with a pulse of green light.  Equipotential surfaces of the density matrix element for the $\ket{0}$ population is plotted according to a rigid-body Euler transformation of the spin-1 operators \cite{PhysRevA.78.033418}.  Note that the $S_z$ term is completely refocused by the echo sequence, while the $S_z^2$ operator evolves, resulting in a different magnitude between the first and sixth states.}
\label{fig2}
\end{center}
\end{figure}

As seen in Eq. \eqref{hint}, the term proportional to $S_z^2$ is sensitive to the frequency drift ($\delta\nu$), whereas the term proportional to $S_z$ varies with $\vec{B}$.  In addition, hyperfine interactions with nearby nuclear spins have a secular correction term to the Hamiltonian $\propto A_{\parallel} S_z I_z$.  By modifying the flip-angle of the first and last pulses of an echo sequence from $\pi/2$ to $\pi/4$, as shown in Fig. \ref{fig2}, the phase accumulation due to the $S_z$ term cancels while that from the $S_z^2$ terms add over the duration $2T$.  

For simplicity, we monitor the eigenkets of $S_z$ to describe the clock, assuming perfect microwave pulses and no optical illumination.  First, optical pumping of the NV prepares the initial state $\ket{\psi_0} = \ket{m_s = 0}$.  Evolution under our drift echo sequence, $U_{echo}$, gives:
\begin{equation}
\ket{\psi_f} = U_{echo} \ket{\psi_0} = \frac{1}{\sqrt{2}} \sin(\phi) \ket{0} - \frac{1}{2}\cos(\phi)\big(\ket{+1} + \ket{-1}\big)
\end{equation}
where $\phi = (D_{gs} - \omega )T = \delta\omega T$.

\begin{figure}[htbp]
\begin{center}
\includegraphics[scale=.40]{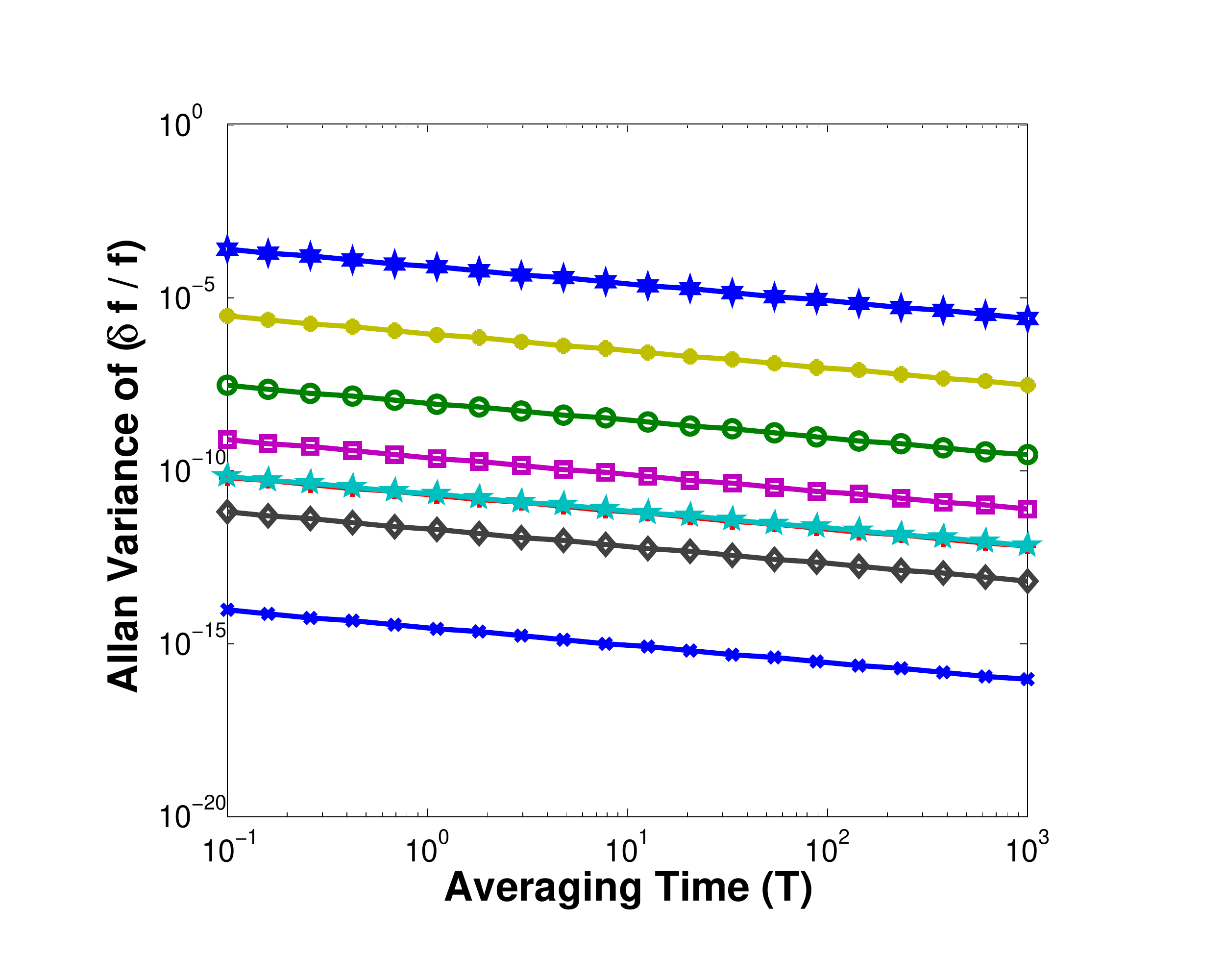}
\caption[Allan Deviation for Atomic and Solid-state standards.]{Allan Deviation for Atomic and Solid-state standards.  Note that the commercial Rb and ensemble NV systems follow the same trend.  Note that in moving from a CW scheme with a single NV center to an ensemble of centers with pulsed excitation and detection, we gain almost six decades of improvement. Legend:  $\times$: Al ion clock,$\circ$: single NVC, $+$: ensemble NVC (0.01pbb x 1mm$^3$),$\star$: commercial Rb clock,$\Box$: Rb chip-clock, $\ast$:surface acoustic wave oscillator, $\diamond$:ovenized quartz reference, \ding{86}: single NVC, steady-state, $\cdot$: ensemble of NVC, steady-state}
\label{fig3}
\end{center}
\end{figure}

Unlike the steady-state case, the NVC is prepared in a state used for time keeping, then measured optically, without simultaneous microwave excitation.  The transient PL response of the NV center can be modeled using projective measurements.  The operator $\hat{M}$ describes the spin expectation value for a PL measurement \cite{Meriles:JCP:2010}, $\hat{M} = a \ket{0}\bra{0} + b \big( \ket{+1}\bra{+1} + \ket{-1}\bra{-1} \big)$, where $a$ and $b$ are independent Poisson random variables. 
The fractional frequency deviation varies as the quantum observable, $\hat{M}$ for our state $\ket{\psi_f}$ according to:
\[
\frac{\delta\omega}{\omega_0} = \frac{1}{\omega_0} \frac{\langle\Delta\hat{M}\rangle}{\vert \partial \langle \hat{M} \rangle/\partial \omega \vert}
\]
where $\langle\Delta\hat{M}^2\rangle = \langle \hat{M}^2 \rangle - \langle \hat{M} \rangle^2$ is the variance of the operator.  For room temperature spin readout of the NVC, $2a = 3b$.  By calculating the moments of $\hat{M}$ and assuming we accumulate the results for a total time $\tau = M^\prime T$, we arrive at:
\[
\Big\langle \frac{\delta\omega}{\omega_0} \Big\rangle_{M^\prime} = \frac{\xi}{ D_{gs} \sqrt{T \tau}}
\]
with $\xi \approx 5$ due to a combination of imperfect spin readout (i.e. $b\neq0$), inefficient collection of photons (i.e $a \ll 1$), and a small ratio of shelving state to radiative lifetime ($\lambda/\gamma$).  Near the onset of  information loss $T=T_2 \approx 1$ms.  With $D_{gs} = 2870$ MHz, this gives a deviation of $\delta\omega/\omega_0 = 8.8 \times 10^{-9} / \sqrt{\tau}$ for a \textit{single} NVC.  Thus the pulsed NV frequency standard dramatically outperforms the CW scheme.  A comparison of the two schemes for different defect concentrations, along with canonical standards, is shown in Figure \ref{fig3}.

The stability for a single NV center can be scaled by collecting the fluorescence of an ensemble of $N$ non-interacting NV centers.  A central limit approach to $S/N$ will improve the stability by a factor of $1/\sqrt{N}$.  To make a reasonable estimate of $N$, we start with the density of pure diamond: $1.74\times10^{23}$ C atoms/cm$^3$.  At an NV$^{-}$ defect fraction of $10^{-11}$ (0.01 part per billion (ppb) ), the density of defects is 1.74 / $\mu$ m$^3$, consistent with commercially available samples.  Thus, assuming scaling the standard to an ensemble a 1 mm$^3$ sample would yield a fractional frequency deviation of $2\times 10^{-13}/\sqrt{\tau}$ for the pulsed scheme and $2\times 10^{-9}/\sqrt{\tau}$ for continuous monitoring.  A diamond film of 1mm$^{2}$ area, 100 um thick, as depicted in Fig. \ref{fig4}, gives $\sigma_y^{pulsed} \sim 2\times 10^{-12}/\sqrt{\tau}$.  Increasing the defect density to 1 ppb would decrease  all values by a another order or magnitude, but dephasing effects may start to arise due to paramagnetic impurities associated with other defect centers \cite{TaylorCappellaro:Mag}.

Several deleterious factors may factor into the operation of the NVC clock to further limit the performance.

\textit{Dephasing.}
The coherent evolution of the NVC is limited by interactions with the environment, such that the $T^{-1/2}$ behavior cannot be maintained for all times.  Unlike an atomic vapor, where collisions and stray magnetic fields dominate the process, the NVC exists in a solid-state lattice of paramagnetic impurities and spin-1/2 carbon nuclei.  Excellent material growth \cite{Balasubramanian:NMat} and precision implantation \cite{Spinicelli:NJP:implant,Toyli:2010zr} can yield crystals free of paramagnetic impurities and provide sufficiently long dephasing times ($T_2$).  For a single NVC in an ultrapure sample, observable spin coherence after an echo sequence of duration $2T$ is reduced by a factor $e^{-(2T/T_2)^n}$, where exponent $n=3$ due to the nuclear spin bath \cite{Slichter}.  In an ensemble of NVC, where each center has a different bath, and thus a different $T_2$, the net effect is a different time constant, $T_e$, with exponent power between $n=2$ and $n=1$ depending on the NVC density \cite{Stanwix:prb}.  We note that the NVC defect density in the sample is a fraction of the total number of substitutional nitrogen impurities given in the sample.  Thus, a heavily doped sample, while improving the $S/N$, can reduce the overall coherence time $T_e$.  Based on the estimates in Ref. \cite{TaylorCappellaro:Mag}, with an NVC density of $10^{17}$/cm$^{3}$, a marginal modification can be expected.  Figure \ref{fig3} shows the Allan deviation floor due to such dephasing considerations.

\textit{Scaling imperfections in ensembles.}
The pulsed NV frequency standard shows a formidable Allan deviation, compared to similarly sized atomic standards, by taking advantage of a spin echo.  Scaling to $N^\prime$ non-interacting defect centers is the straightforward way to reduce the deviation by $\sqrt{N^\prime}$. but, several deleterious effects may arise.  First, the local strain fields within the crystal may vary from center to center due to inhomogeneous broadening.  We can estimate the effects of strain by assuming a large enough number of centers such that $\Pi_z$ fields given by a zero mean Gaussian distribution with variance $\Sigma^2$.  In the limit that this inhomogeneous broadening linewidth exceeds the homogeneous broadening, which is responsible for the $T_2$ time of the coherence, we would further limit the sensing time by a new value $T_2^*$.  We estimate a $\Sigma \approx 10^{6}$ V/cm, 
yielding a $T_2^*$ of $\approx  30$ns, according to the relationship  $(T_2^*)^{-2} = d_\parallel^2 \Sigma^2/2$.  This would reduce the stability by more than two orders of magnitude.  Modern CVD samples and annealing techniques may be able to limit variations of local strain further.  In addition, it is known that the resonance frequency corresponding to the zero-field splitting of the NV$^{-}$ center has a temperature dependence $d\Delta/dT = -74.2(7)$ kHz/K \cite{AcostaTempPhysRevLett.104.070801}.  This could be due to lattice expansion and an associated strain term that shifts the local strain of each NVC, or it could be to a relaxation of the NV $C_{3v}$ symmetry with temperature dependence.  Such temperature dependence makes this passive standard behave like a crystal oscillator and poses a challenge for maintaining a precise operating temperature.  It is not clear how other effects common to crystal oscillators, e.g. aging, may be affected by strain relief over a period of years.
%

Looking forward, the diamond frequency standard promises a fully chip-integratable solid-state time-keeping platform whose performance rivals that of modern atomic clocks, both laboratory and chip-scale.  Such miniaturized clocks can be easily integrated into scientific and consumer electronics and could be used in harsh external environments.  Devices benefiting include those for wireless communication and GPS navigation with improved tolerance to jamming \cite{Vig1993}. In addition, the clock's center-frequency of 2.87 GHz would enable wide-bandwidth date rate in next-generation cellular communications ($f \ge 40$ GHz).  At these frequencies, phase-sensitive data encoding is limited by the phase noise of the derived frequency source, which is usually a MHz range mechanical oscillator frequency upconverted by a factor N $\ge 1000$.  The low noise spectrum of the mechanical oscillator, scales with $N^2$, thus degrading performance.  A low-power, portable, and stable on-board oscillator in the GHz range can avoid this quadratic stability loss (Fig. \ref{fig3}).

\section{Conclusions}
In summary, we have proposed a solid-state spin-optical frequency standard based on the NV defect center in diamond.  Due to the center's relatively long lifetime, high density of spins, and optical detection, we estimate a time stability exceeding $\sigma_y = 2\times10^{-12} \tau^{-1/2}$, rivaling that of the newest chip-scale Cs and Rb standards.  Many of the technological advances, such as the surface emitting lasers and chip-scale detectors, are readily deployable to this system, with the added benefit of the atomic system being inside the diamond substrate.  We anticipate the ability to improve the frequency standard using repetitive readout \cite{Jiang} or entanglement assisted \cite{Neumann} schemes, by using IR absorption detection \cite{AcostaPRB2010}, and by deploying nanophotonic \cite{barclay:191115}
advances in diamond to further improve the sensitivity of this device.

\section{Acknowledgments}

The authors acknowledge Mikhail Lukin, Colm Ryan, Peter Kinget, and Tanya Zelevinsky for useful discussions.  This work was supported by the U.S. Air Force Office of Scientific Research Young Investigator Program, AFOSR Grant No. FA9550-11-1-0014, supervised by Dr. Gernot Pomrenke.
\bibliography{biblio.bib}

\end{document}